\documentclass[11pt]{article}
\usepackage{fullpage}
\usepackage{algorithmic}
\usepackage{algorithm}
\usepackage{array}
\usepackage{amsthm}
\usepackage{amssymb}
\usepackage{amsmath}
\usepackage{amsfonts}
\usepackage{booktabs}
\usepackage{bbm}
\usepackage{epsfig}
\usepackage{epstopdf}
\usepackage{enumerate}
\usepackage{eufrak}
\usepackage{graphicx}
\usepackage{latexsym}
\usepackage{mathrsfs}
\usepackage{multirow}

\usepackage[symbol]{footmisc}

\usepackage{bigstrut,multirow}

\theoremstyle{remark}

\theoremstyle{definition}

\newtheorem*{mechanism*}{Mechanism}

\makeatletter
\newcommand\figcaption{\def\@captype{figure}\caption}
\newcommand\tabcaption{\def\@captype{table}\caption}
\makeatother

\DeclareMathAlphabet{\mathpzc}{OT1}{pzc}{m}{it}

\begin{document}
\allowdisplaybreaks[4]

\title{Mechanism Design for Facility Location Problems: A Survey}

\date{}
\maketitle
\vspace{-3em}

\begin{center}
\author{
Hau Chan$^1$~~
\and
Aris Filos-Ratsikas$^2$~~
\and
Bo Li$^{3}$~~
\and
Minming Li$^4$~~
\and
Chenhao Wang$^1$\footnote{Corresponding author}\\
${}$\\
$^1$Department of Computer Science and Engineering, University of Nebraska-Lincoln\\
$^2$Department of Computer Science, University of Liverpool\\
$^3$Department of Computing, The Hong Kong Polytechnic University\\
$^4$Department of Computer Science, City University of Hong Kong\\
\medskip
hchan3@unl.edu,
aris.filos-ratsikas@liverpool.ac.uk,
comp-bo.li@polyu.edu.hk,
minming.li@cityu.edu.hk,
chenhwang4-c@my.cityu.edu.hk
}
\end{center}

\pagestyle{plain}

\vspace{1em}

\begin{abstract}
The study of \emph{approximate mechanism design} for facility location problems has been in the center of research at the intersection of artificial intelligence and economics for the last decades, largely due to its practical importance in various domains, such as social planning and clustering.
At a high level, the goal is to design mechanisms to select a set of locations on which to build a set of  \emph{facilities}, aiming to optimize some social objective and ensure desirable properties based on the preferences of \emph{strategic agents}, who might have incentives to misreport their private information such as their locations.
This paper presents a comprehensive survey of the significant progress that has been made since the introduction of the problem, highlighting the different variants and methodologies, as well as the most interesting directions for future research.
\end{abstract}

\section{Introduction}


The study of facility location problems arises from combinatorial optimization, and aims at computing the optimal placement of facilities to minimize transportation costs for servicing customers \cite{hochbaum1982heuristics}. Besides locating actual facilities, the problem has also found a wide range of applications in other fields such as healthcare \cite{ahmadi2017survey}, clustering \cite{friedman2001elements}, and even problems that are not of a geographical nature, ranging from as simple as choosing the temperature for a classroom, to more advanced ones, like selecting a committee to represent people with different political views. Because of its practical importance, variants of facility location problems have attracted significant attention for a long time from different fields, such as operations research,
theoretical computer science,
economics
and computational game theory.


In the last two research areas mentioned above, facility location problems are studied under a different light: strategic customers (referred to as \emph{agents}) may have incentives to misreport their private information, such as their preferred locations, in order to manipulate the final outcome in their favor. A \emph{strategyproof mechanism} is a predefined rule which maps the preferences of the agents into appropriately chosen locations of the facilities, in a way that does not incentivize the agents to engage in such strategic considerations. Strategyproof mechanisms are certainly desirable, but they come at a cost; to ensure the truthful behavior of the participants, they often need to return sub-optimal solutions. One of the main investigations of the field of \emph{computational social choice theory} \cite{brandt2016handbook} is precisely to quantify this loss in efficiency due to the requirement for strategyproofness.

In fact, more than a decade ago, Procaccia and Tennenholtz \cite{procaccia2009approximate} used one of the most basic variants of facility location problems to put forward the agenda of \emph{approximate mechanism design without money}, which advocates the study of strategyproof mechanisms for various optimization problems through the lens of the \emph{approximation ratio}, a celebrated notion in the field of theoretical computer science and approximation algorithms. 
The major difference between this agenda and prior work in algorithms is that here, the need for approximation does not come from computational limitations, but primarily due to the need for strategyproofness. The setting studied by Procaccia and Tennenholtz \cite{procaccia2009approximate} is that of a single facility to be built on a real line, and in which the agents' preferences are given by their ideal locations (their ``peaks''), together with linear cost functions that are increasing at a fixed rate as one moves away from the peak.

A key characteristic of this setting, which makes it particularly amenable to the design of strategyproof mechanisms, is that the preferences of the agents are \emph{single-peaked}.\footnote{In fact, these are specific types of single-peaked preferences called \emph{$1$-Euclidean}, see \cite{hotelling1990stability}.} 
This type of preferences, originally studied by Black \cite{black1948rationale}, is known to escape classic impossibility results 
and allows for winners of pairwise majority votes (referred to in the literature as ``Condorcet winners''). Single-peaked preferences were popularized by the result of Moulin \cite{moulin1980strategy}, who identified all the possible strategyproof mechanisms for this setting by means of a \emph{characterization}. Among those mechanisms, we highlight the \emph{Median-point mechanism} below, phrased in appropriate terminology for facility location problems.

\begin{mechanism*}[Median-point mechanism]
Place the single facility at the  median of the peaks of the agents.
\end{mechanism*}
\noindent It is not hard to see that this mechanism is strategyproof; an agent can only affect the position of the median by reporting her peak to be on the ``opposite'' side of the median, but in that case the median can only ``move away'' from her peak. As a matter of fact, the Median-point mechanism is simply a majority outcome, something that was already observed by Black \cite{black1948rationale} in his original paper.

Since the introduction of the agenda by Procaccia and Tennenholtz \cite{procaccia2009approximate}, research in this area has flourished. For facility location problems in particular, a very fruitful line of work in the major venues in artificial intelligence and economics and computation has been concerned with several variants and generalizations of the main setting, often migrating from the ``convenience'' of single-peakedness and thus giving rise to much more challenging problems. The ultimate goal of this survey is to provide a comprehensive overview of the achievements of this literature on designing strategyproof mechanisms for these different variants. The only other survey on strategyproof facility location problems that we are aware of is that of \cite{cheng2015survey}, which is however mainly focused on the more classic settings. While we do highlight some of the earlier results on the problem as well, we put extra emphasis on the results of the last half decade, which has witnessed remarkable achievements.

\subsection{The Classic Setting}
We formally introduce the classic facility location setting in this section, which was studied in most of the earlier works on the problem; various extensions are presented in later sections.
Let $N=\{1,\ldots,n\}$ be a set of agents who are located in a metric space $(M,d)$, where $M$ is a set of all possible positions and $d:M\times M\rightarrow \mathbb R$ is a distance function that satisfies the triangle inequality. We use $\mathbf x=(x_1,\ldots,x_n)\in M^n$ to denote the location profile of the $n$ agents.
In a $k$-facility location setting, where the goal is to place $k$ facilities, a (deterministic) \emph{mechanism} is a function $f:M^n\rightarrow M^k$ that maps the agents' locations $\mathbf x$ to $k$ locations of the facilities.
A randomized mechanism outputs a probability distribution over all subsets of $M$ with size $k$.
Given a location profile $f(\mathbf x)$ of $k$ facilities, each agent $i\in N$ has a cost $c(f(\mathbf x),x_i)$.
Usually, the cost is defined as the distance to the nearest facility, i.e., $c(f(\mathbf x),x_i)=\min_{y\in f(\mathbf x)}d(x_i,y)$.
When $f(\mathbf x)$ is a distribution, the cost is defined as the expected distance. We remark that when there is only one facility ($k=1$) and the metric space is the real line, we recover the original setting of \cite{procaccia2009approximate}.

In the strategic setting, the location $x_i$ of each agent $i\in N$ is private information.
For any $S\subseteq N$ and $i\in N$, we use $\mathbf x_S$ to denote $\mathbf x$ projected to $S$ and $\mathbf x_{-i}$ to $N\backslash\{i\}$.
Therefore, mechanism $f$ is {\em strategyproof} if for all $\mathbf x\in M^n$,
\[
c(f(\mathbf x),x_i)\le c(f(x_i,\mathbf x_{-i}),x_i) \text{ for all $i\in N,x_i'\in M$}.
\]
In simple words, a strategyproof mechanism requires that no agent can benefit from misreporting, regardless of the reported positions of the others agents.
A stronger notion is \emph{group strategyproofness}, which requires that for any location profile $\mathbf x$ and any coalition $S\subseteq N$, there is no joint deviation $\mathbf x_S'$
 of the agents in $S$ such that all the agents in $S$ benefit.

As we mentioned earlier, the goal of approximate mechanism design without money is to optimize some objective under the requirement that the employed mechanisms are strategyproof. The two most widely studied objectives in the literature of the problem are the social cost and the maximum cost. Formally, given agent profile $\mathbf x$ and facility locations $\mathbf y$, the \emph{social cost} is defined as
\[
SC(\mathbf y,\mathbf x)=\sum_{i\in N}c_i(\mathbf y,x_i)
\]
and the \emph{maximum cost} is defined as
\[
MC(\mathbf y,\mathbf x)=\max_{i\in N}c_i(\mathbf y,x_i).
\]
In economic terms, the social cost is analogous to the \emph{utilitarian social welfare}, whereas the maximum cost is analogous to the \emph{egalitarian social welfare}.

The objective of a strategyproof mechanism $f$ is to compute locations $\mathbf y$ such that $SC(\mathbf y,\mathbf x)$ or $MC(\mathbf y,\mathbf x)$ is minimized. A mechanism $f$ has an approximation ratio $\alpha\ge 1$ for the social cost objective, if for any location profile $\mathbf x\in M^n$,
\[
SC(f(\mathbf x),\mathbf x)\le \alpha\cdot \min_{\mathbf y\in M^k}SC(\mathbf y,\mathbf x).
\]
The approximation ratio is defined similarly for the maximum cost objective, or any of the other objectives that we will discuss in later sections.


\medskip\noindent\textbf{Challenges.} Strategyproofness is a rather stringent and fragile property, which makes the design of good strategyproof mechanisms particularly challenging, since robustness to strategic behavior has to be guaranteed for \emph{every possible instance of the problem}. To demonstrate this, consider the single-facility on the real line setting studied by Procaccia and Tennenholtz \cite{procaccia2009approximate}. It is not hard to observe that the Median-point mechanism is in fact optimal for the social cost objective; for any other choice, the cost of more than half of the agents is increased by the same amount that the cost of the remaining (less than half) of the agents is decreased by. For the maximum cost objective however, the optimal choice, which is to place the facility at the midpoint of the two extreme reported locations, is no longer strategyproof; this is due to the fact e.g., the agent at the leftmost position would be incentivized to report a position further to the left, forcing the midpoint to be closer to her real location.
The Median-point mechanism in this case only achieves a $2$-approximation and in fact, this is the best possible among any strategyproof mechanism for the problem \cite{procaccia2009approximate}.
In more general metric spaces and for more facilities, the situation becomes ever more challenging.

One way to circumvent such barriers is to use randomization, and consider a slightly weaker notion of strategyproofness, often referred to as \emph{truthfulness in expectation}; this notion stipulates that an agent will not decrease her expected cost by misreporting, even if her cost might in fact decrease in some realizations of randomness. An example of such a mechanism proposed by Procaccia and Tennenholtz \cite{procaccia2009approximate} for their setting is the following one.

\begin{mechanism*}[Left-right-middle (LRM) mechanism]
Select the peak of the leftmost agent with probability $1/4$, the peak of the rightmost agent with probability $1/4$ and the midpoint of the interval defined by those peaks with probability $1/2$.
\end{mechanism*}
It was shown in \cite{procaccia2009approximate} that this mechanism achieves an approximation ratio of $3/2$ for the maximum cost objective, a notable improvement of the $2$-approximation barrier for deterministic mechanisms. In more general settings, the benefits of randomization are far more pronounced, improving the best possible approximations from linear functions in the number of agents to small constants \cite{lu2010asymptotically}.



The ultimate goal of mechanism design for facility location is to be applied to practical scenarios of social planning, by providing strategyproof mechanisms with guarantees on social efficiency. While the classic setting described above is a crucial starting point in these investigations, the plethora of different variants that have appeared over the years take further steps towards applicability in real-world scenarios.

\medskip\noindent\textbf{Roadmap.} We first introduce  the results on the classic setting of strategyproof facility location in Section~\ref{sec:class}, and we proceed to discuss more recent variants in the next sections. Section~\ref{sec:agent} is devoted to settings with more advanced agents' preferences, whereas Section~\ref{sec:mech_more_involved} is concerned with mechanisms that either have enhanced capabilities, or need to be robust against stronger types of manipulations.  In Section~\ref{sec:fac} we discuss settings with more complicated constraints on the facilities, and in Section \ref{sec:mec}, we discuss several further variants of the problem along different axes. In Section~\ref{sec:future}, we identify interesting avenues for future work on the problem.


\section{The Classic Facility Location Setting: Methods and Results}\label{sec:class}
We begin our exposition with the earlier works on the problem, which we refer to as the ``classic'' setting.

\subsection{Locating a Single Facility}
The foundations of locating a single facility on the real line are based on the work of \cite{moulin1980strategy}. As we explained earlier, the Median-point mechanism is strategyproof (in fact, group strategyproof) and optimal for the social cost objective, and is $2$-approximate for the maximum cost, which is the best possible among strategyproof mechanisms \cite{procaccia2009approximate}. In terms of randomized mechanisms, the LRM mechanism is $3/2$-approximate, which is also best possible among randomized strategyproof mechanisms, as proven in the same work. In that sense, the \emph{approximate} mechanism design problem for a single facility on the line was settled in \cite{procaccia2009approximate}.


The single-facility problem for more general metrics was studied by Alon \emph{et al.} \cite{alon2010strategyproof}, who designed strategyproof mechanisms for circles and general graphs. A highlight of their results is that randomization can be used to escape a rather strong impossibility result, namely that in circle networks, the approximation ratio of any deterministic mechanism for the social cost is $\Omega(n)$. In particular, they employed the following well-known randomized strategyproof mechanism and showed that it achieves an approximation ratio of $2-2/n$ on general graphs.

\begin{mechanism*}(Random Dictatorship (RD))
Select an agent uniformly at random, and place the facility at her peak.
\end{mechanism*}

\noindent Meir \cite{DBLP:conf/sagt/Meir19} showed that for three agents on a circle, improvements over the RD mechanism are possible. Feldman and Wilf \cite{feldman2013strategyproof} proposed a family of generalized mechanisms on trees (the ``parameterized boomerang mechanisms'') which generalize both the RD and the Median-point mechanism. Dokow \emph{et al.} \cite{dokow2012mechanism} presented similar characterizations to that of \cite{schummer2002strategy} for discrete lines and circles. Recently, Tang \emph{et al.} \cite{tang2020characterization} provided a characterization of deterministic and randomized group strategyproof unanimous mechanisms in convex spaces (generalizing and extending some classic results of \cite{bordes1990strategy}),
and showed approximation ratio bounds that can be obtained as a consequence of their characterization. Filimonov and Meir \cite{filimonov2021strategyproof} gave a full characterization of onto and strategyproof mechanisms in discrete trees.

\subsection{Locating Two Facilities}
For two facilities, most of the related work is focused on the line metric. This setting was in fact also studied by Procaccia and Tennenholtz  \cite{procaccia2009approximate} in their original paper. In particular, they showed that the very simple mechanism that places one facility on the leftmost peak and the other on the rightmost peak is group strategyproof and achieves an $(n-2)$-~approximation for the social cost. They also provided a lower bound of $3/2$ for any strategyproof mechanism; this was later improved to almost $2$  \cite{lu2009tighter} and to an asymptotically tight bound of $\Omega(n)$ \cite{lu2010asymptotically}.  Fotakis and Tzamos \cite{fotakis2014power} proved a tight lower bound of $n-2$, thus settling the social cost objective for this setting.
These works also considered randomized strategyproof mechanisms, with the best known approximation ratio being $4$, given by the {\em Proportional Mechanism} of \cite{lu2010asymptotically}. The same approximation ratio in fact applies to general metrics as well.


For the maximum cost objective, Procaccia and Tennenholtz  \cite{procaccia2009approximate} showed that their ``two extremes'' mechanism achieves a best possible $2$-approximation, and showed how randomization can be used to achieve improved bounds.

\subsection{Beyond Two Facilities}
Escoffier \emph{et al.} \cite{DBLP:conf/aldt/EscoffierGTPS11} designed strategyproof mechanisms to locate $n-1$ facilities in both general metric spaces and trees, providing linear upper bounds and constant lower bounds.
For the real line metric, Fotakis and Tzamos \cite{fotakis2016strategyproof} presented a randomized strategyproof mechanism for any number of facilities $k$, the {\em Equal Cost} mechanism, which achieves approximation ratios of $n$ and $2$ for the social cost and the maximum cost respectively. Interestingly, the mechanism achieves these bounds even in the more general case where the cost functions are concave. 
For the social cost objective, Fotakis and Tzamos \cite{fotakis2014power} showed that \emph{anonymous} deterministic strategyproof mechanisms for $k\leq 3$ cannot achieve any bounded ratio, even if the metric is the real line and there are $3$ facilities and $4$ agents; this includes the \emph{Percentile mechanisms} studied in  \cite{sui2013analysis}. Walsh \cite{walsh2020strategy1} studied the $k$-facility location problem in $2$-dimensional Euclidean and Manhattan spaces and provided impossibility results in terms of axiomatic properties.



\section{More Involved Preferences}\label{sec:agent}
A common characteristic of the earlier works in facility location is that they study settings in which agents have a most-preferred point, and the cost functions are given as distances from that point. The second half of the past decade however witnessed the emergence of a plethora of results on different models, in which the preferences of the agents are more involved, giving rise to new challenges in terms of the design of good strategyproof mechanisms. In this section, we highlight the most prominent of those variants.

\subsection{Obnoxious Facility Location} The location of an \emph{obnoxious} (in the sense of ``unwanted'') facility is in fact a rather classic problem in the literature on algorithms and optimization (e.g., see \cite{church1978locating}). In this setting each agent would rather have the facility (e.g., a waste facility) built as far away as possible from her specified location (e.g., the location of her house); and the distance from that location to the facility is interpreted as her \emph{utility}. In the context of approximate mechanism design, this setting was studied most notably by Cheng \emph{et al.} \cite{cheng2013strategy}, who proposed strategyproof mechanisms for networks, which are based on a \emph{majority vote} between two prespecified locations. For the line metric and different objectives, the problem was studied by Ye \emph{et al.} \cite{ye2015strategy}.  Ibara and Nagamochi \cite{ibara2012characterizing} provided characterization results for (group) strategyproof mechanisms for some special cases of the problem.

\subsection{Heterogeneous Facility Location}

Building upon the ideas of the obnoxious facility literature, a follow-up line of work, (e.g., \cite{feigenbaum2015strategyproof,zou2015facility,paolo2014heterogeneous,paolo2015heterogeneous}) considered models in which facilities can be both \emph{desirable} and \emph{undesirable}, from the perspective of \emph{different} agents. In terms of strategyproofness, here it makes sense to distinguish between the cases where the \emph{preferences} (desirable vs undesirable), or the locations of the agents are private information, or both.
While these settings appear under many different names in the literature, we will use the term \emph{heterogeneous facility location} to classify them under the same umbrella; we highlight the major variants within this part of the literature.

\medskip \noindent \textbf{Dual preferences.}
One of the first heterogeneous facility location models studied in this literature was the \emph{dual preference model}, proposed independently by \cite{zou2015facility} and \cite{feigenbaum2015strategyproof}. In this setting each agent finds some facility either desirable or undesirable, but different agents have potentially different opinions. For the case of known locations, Zou and Li \cite{zou2015facility} proved that it is possible to design an optimal strategyproof mechanism for the \emph{social welfare} (the equivalent of the social cost for settings with utilities), whereas when both preferences and locations are unknown,  they proposed a $3$-approximate deterministic group strategyproof mechanism. In fact, the results of \cite{ibara2012characterizing} for the obnoxious facility location (which is a special case of the dual preference setting) imply that this is actually best possible among strategyproof mechanisms; this was also proven independently by Feigenbaum and Sethuraman \cite{feigenbaum2015strategyproof}, who further provided lower bounds for randomized mechanisms. Kyropoulouetal  \cite{kyropoulou2019mechanism} studied a related setting in which the facility can be located on a constrained feasible region of the Euclidean plane.

\medskip \noindent \textbf{Optional preferences.}
A somewhat different model was considered first by Serafino and Ventre \cite{paolo2014heterogeneous,paolo2015heterogeneous}, that of \emph{optional preferences}. In that setting, there are two facilities to be built, each agent is ``interested'' in either facility or both, and her cost is equal to the sum of distances from her position to the locations of the facilities she is interested in (the \emph{sum} function). The setting was later extended by Yuan \emph{et al.} \cite{yuan2016} who considered the \emph{min} and \emph{max} functions instead, and they proved upper and lower bounds on the approximation ratio of deterministic strategyproof mechanisms for different objective functions. The upper bound for the min function was later improved \cite{li2019strategyproof}.  Recently, Deligkas \emph{et al.} \cite{deligkas2021heterogeneous}  studied a utility version: A mechanism takes as input the
positions and the preferences of the agents, and chooses to locate a single facility.  Each agent has a utility equal to one minus her distance to the facility located if she is interested in it, and zero otherwise.  Under three different settings depending
on the level of agent-related information that is public or private, they designed strategyproof mechanisms that achieve a good approximation  for maximizing the total utility.

\medskip \noindent  \textbf{Heterogeneous preferences for $k$ facilities.}
Anastasiadis and Deligkas \cite{anastasiadis2018heterogeneous} considered a setting with $k$ facilities such that, for each facility, an agent prefers to be either close to the facility, far from it, or is indifferent between the two choices. In a sense, their setting merges the dual preference and the optional preference models described above, and extends them to more facilities. They studied mainly the egalitarian welfare (the utility equivalent of the maximum cost objective), but also obtained some results on the social welfare as a byproduct.


\medskip \noindent \textbf{Fractional preferences.}
Another generalization of the optional preference model was proposed in
\cite{fong2018facility}, that of the \emph{fractional preference model}. In this setting,
the preference of each agent for the facility is a number between $0$ and $1$ (rather than either $0$ or $1$, as in the case of optional preferences), with these numbers summing up to $1$. Among other results, the authors show that when the locations are known, it is possible to achieve the maximum social welfare by a strategyproof mechanism, when the preferences are known, a $2$ approximation is possible and when the locations and the preferences are private, there is a deterministic $4$-approximate mechanism for the problem.




\section{More Involved Incentives}\label{sec:mech_more_involved}
In all the aforementioned works, the ability of the agents to misreport is limited to either their locations or their preferences. In this section we discuss some representative variants where the space of possible manipulations is extended, as well as variants in which mechanisms are given additional power to deal with the agents' incentives.

\medskip \noindent \textbf{Multiple locations.}
Besides the basic setting, Procaccia and Tennenholtz \cite{procaccia2009approximate} in their seminal paper also considered a natural extension where each agent controls \emph{multiple locations}, and her cost is either the total distance or the maximum distance from her locations to the facility. Hossain \emph{et al.} \cite{hossain2020surprising} further assumed that each agent may control multiple locations with different degrees of importance, and designed mechanisms which
elicit the locations of the agents, as well as different levels of information about their importance. In their setting, in addition to misreporting their locations, agents may also \emph{hide} some of their locations, if that serves their purposes. Yan and Chen \cite{yan2020strategyproof} enhanced the space of possible manipulations even further, assuming that the agents can \emph{replicate} some of their locations as well, and fully characterized all mechanisms that are anonymous, efficient, and robust to manipulation with respect to this richer space of strategic behavior.
 Babaioff \emph{et al.} \cite{babaioff2016mechanism} studied three-stage mechanisms by introducing \emph{strategic mediators}, where each agent is associated with exactly one
mediator, and the cost of a mediator is the total cost of the agents
that she represents. In this setting, the agents first strategically report their locations to the mediators, and then the mediators strategically report them to the mechanism.


\medskip \noindent \textbf{False-name manipulations.}
Another type of manipulation which has been studied in the literature of mechanism design for \emph{anonymous} environments is that of \emph{false-name manipulation} (also known as \emph{sybil-proofness}, e.g., see \cite{babaioff2012bitcoin}), where an agent can report more than once by creating and using fake identifiers (e.g., different email addresses or repeatedly
logging onto an online service).  A mechanism that is robust against standard and false-name manipulations is called \emph{false-name-proof} (FNP), and is thus \emph{stronger} than a simply strategyproof mechanism. False-name manipulations in facility location were first considered by Todo \emph{et al.} \cite{DBLP:conf/atal/TodoIY11}, who provided a characterization of FNP mechanisms for single-facility location on the real line, and proved that the mechanism that places the facility at the location of the leftmost agent achieves the best possible approximation for both the social cost and the maximum cost. Sonoda \emph{et al.} \cite{sonoda2016false} studied FNP mechanisms for locating two facilities and proved that the ``two extreme'' mechanism is the best possible deterministic FNP mechanism on a line, while randomized mechanisms can achieve improved approximations. 
Nehama \emph{et al.} \cite{nehama2019manipulations} defined a family of graphs that is called ZV-line graphs, and proved that there is a general facility location mechanism for these graphs that is strategyproof, false-name-proof and Pareto optimal.
Other works that studied this type of manipulation in the context of facility location are \cite{ono2017rename,todo2020false}.


\medskip \noindent \textbf{Mechanism design with verification.}
Motivated by settings in which strategyproofness is not enough to guarantee small approximation ratios, Fotakis and Tzamos
\cite{fotakis2010winner} and  Nissim \emph{et al.} \cite{NissimST12} first considered \emph{winner-imposing mechanisms}, which are capable of penalizing possibly deceitful agents by restricting how these agents interact with the outcome, e.g., by requiring agents to ``use'' the facility that is closer to their \emph{reported} location. Fotakis and Tzamos \cite{fotakis2010winner} proved that a winner-imposing extension of the Proportional Mechanism of \cite{lu2010asymptotically} is strategyproof and achieves a $4k$-approximation for the $k$-Facility Location game on the real line.
In general, this line of research fits under the umbrella of mechanism design with verification (e.g., see \cite{sigecom/CaragiannisESY12}). 


\section{Constrained Facilities}\label{sec:fac}

Besides the different preference structures that were studied in previous models, the motivation from real-world applications drove researchers to consider additional constraints on the facilities rather than the agents, which might have to do with their capacities (e.g., in terms of physical storage or service time) or their possible locations. We highlight the major associated results below.

\medskip \noindent \textbf{Limited locations.}
In the standard setting, the location of the facility is assumed to be some point in a continuum of possible points. In reality however, the ``allowed'' points for the placement of the facility could be limited - e.g., a bus stop needs to be located on the main road. Feldman \emph{et al.} \cite{Distortion4} studied the single-facility location setting in the context of voting embedded in a metric space, a topic very much related to that of \emph{distortion in metric social choice} \cite{anshelevich2018approximating}. Among other results, for the line metric, they provided a deterministic 3-approximation mechanism, which places the facility at the candidate location that is closest to the median agent, as well as a randomized $2$-approximation mechanism for the social cost, together with matching lower bounds. Tang \emph{et al.} \cite{tang2020mechanism} further considered the maximum cost objective and the case of two-facility location. Walsh  \cite{walsh2020strategy3} considered a related setting in which each facility can be placed in a finite set of feasible subintervals. 



\medskip \noindent \textbf{Distance constraints.}
Zou and Li \cite{zou2015facility} initiated the study of settings with two facilities, in which the placement of one of them imposes constraints on the placement of the other. In particular, they considered the \emph{maximum distance requirement}, namely that the distance between the two facilities cannot exceed a certain threshold, for the \emph{two-opposite-facility problem}; in this setting, there is one desired and one undesired facility, and the utility of each agent is the distance to undesired one minus the distance to the desired one. 
Later, Tang \emph{et al.} \cite{tang11059mechanism} and Chen \emph{et al.} \cite{chen2021tight} relaxed the maximum distance constraint by instead imposing a \emph{penalty}, which is linear in the degree of violation.
Xu \emph{et al.} \cite{journals/jair/Xu21Two}
studied the analogous \emph{minimum} distance requirement for heterogeneous two-facility locations, where the cost of an agent is the sum of her distances to the two facilities and  Xu \emph{et al.} \cite{xu2020strategyproof} studied the case with an \emph{exact} distance requirement.

\medskip \noindent \textbf{Capacitated facilities.}
The study of \emph{capacitated facility location games} was initiated by Aziz \emph{et al.} \cite{aziz2020capacity}. They studied a single-facility location problem with capacity constraints on the real line and  provided a complete characterization of strategyproof mechanisms, quite similar to that of \cite{moulin1980strategy}. In a follow-up work, Aziz \emph{et al.} \cite{aziz2020facility} considered a more general setting with $k$ facilities and different capacities for each facility. Among other results, the authors proposed the \emph{Extended Endpoint Mechanism} for the case of two facilities, which achieves a tight approximation ratio matching the lower bound proven in \cite{aziz2020capacity}.


\section{Further Variants}\label{sec:mec}
In this section, we present some further interesting variants that do not quite fit any of the categories presented earlier.

\smallskip \noindent \textbf{Other social objectives. }
Besides the social cost and the maximum cost, the literature has also been concerned with other objectives. Feigenbaum \emph{et al.} \cite{Feigenbaum2014} considered the minimization of the $L_p$-norm of costs for the real line metric, whereas Feldman and Wilf \cite{feldman2013strategyproof} studied the objective of minimizing the sum of squares of the costs. In a markedly different direction, Cai \emph{et al.} \cite{DBLP:conf/ijcai/CaiFT16} studied the \emph{minimax envy} objective for single-facility location,
where the envy of an agent with respect to another agent is simply their difference in distance from the facility, and the goal is to minimize the maximum envy over all the agents; we remark that contrary to most of the present results in this survey, the approximations of \cite{DBLP:conf/ijcai/CaiFT16} are additive.
In a similar vain, Ding \emph{et al.} \cite{DingLCFN20} and Liu \emph{et al.} \cite{liu2021multiple} studied the \emph{envy ratio}, which is defined as the maximum over the ratios between any two agents’ utilities. Mei \emph{et al.}
\cite{mei2019facility} introduced a \emph{happiness} factor within $[0,1]$ to measure the
difference between the best possible facility location for an agent and the one selected by the mechanism, in both settings of standard and obnoxious facility games. In this regime, the goal of the agent is to maximize her happiness factor, and the mechanism aims to maximize the sum of those factors.

\smallskip \noindent \textbf{Double-peaked preferences. } Filos-Ratsikas \emph{et al.}
 \cite{DBLP:conf/aaai/Filos-RatsikasL15} studied the single-facility location problem with double-peaked preferences, where each agent has  \emph{two peaks} on the real line, and the cost increases as one moves away from either peak. The authors provided approximation ratio upper and lower bounds for deterministic and randomized strategyproof mechanisms, as well as an axiomatic characterization of group strategyproof mechanisms that satisfy some standard neutrality properties.




\smallskip \noindent \textbf{Externalities.} Li \emph{et al.} \cite{DBLP:conf/atal/LiMXZZ19} considered settings where agents mutually influence each other, i.e., when one agent's utility positively or negatively affect some other agents'.

\smallskip \noindent \textbf{Weighted agents.} Zhang and Li \cite{zhang2014strategyproof} introduced \emph{weights} to the agents, and showed that when weights are private information, the best that a deterministic strategyproof mechanism can do is to simply ignore them.

\smallskip \noindent \textbf{Distributed facility location.} Recently, Filos-Ratsikas and Voudouris \cite{filos2020approximate} introduced a setting in which the facility location is selected as part of a distributed process: first, agents within groups (or \emph{districts}) decide on a representative location and then the mechanism, oblivious to the actual locations of the agents, decides on a location from the set of representatives. The authors proved that the best possible strategyproof mechanism for this setting has an approximation ratio of $3$.

\smallskip \noindent \textbf{Approximation and variance.} Procaccia \emph{et al.} \cite{procaccia2018approximation} considered the topic of reducing the variance of randomized strategyproof mechanisms while maintaining good approximation guarantees, by using the single-facility location problem as the application domain.

 \smallskip \noindent {\textbf{Additive approximations.}
Instead of multiplicative approximation ration, Golomb and Tzamos \cite{golomb2017truthful} studied the worst-case additive approximation, and presented tight bounds for mechanisms locating a single facility in both deterministic and randomized cases.
They also argued that this measure is better suited for some
situations.}

\smallskip \noindent {\textbf{Shuttle facility games.}  Fukui \emph{et al.}
 \cite{fukui2020group} introduced the shuttle facility games, where each agent' location is actually an interval in the line representing her commuting route.
They designed a corresponding group strategyproof mechanism which also achieves the optimal social welfare.
}

\smallskip \noindent {\textbf{Mechanisms with payments.} Archer and Tardos
\cite{archer2001truthful} studied a conceptually similar but fundamentally different facility location problem, a monetary model. Given a set of facilities holders and a set of customers, the facilities holders are strategic players and are required to report their private cost for them to be built. Once receiving the reports, the government needs to select a subset of facilities to build, and monetarily compensate these facilities with payment to guarantee truthful reports. They proven that any algorithm that solves the uncapacitated facility location problem optimally admits a truthful mechanism. Li \emph{et al.} \cite{li2020budgeted} further studied this monetary model  with a sharp budget constraint: the total payment of a mechanism is below a given budget. Chen \emph{et al.} \cite{chen2019truthful} considered a dual-role setting where every agent plays both roles of facility holder and customer. 
}

\smallskip \noindent \textbf{Automated mechanism design.} There are also some works that use machine learning and deep learning approaches to design automated mechanisms; for example, see \cite{DBLP:conf/ijcai/GolowichNP18,DBLP:conf/ijcai/Narasimhan0P16}. 


\section{Future Research Directions}\label{sec:future}
Many open problems related to all the different facility location variants were already implicit in our exposition in previous sections; we conclude this survey by identifying some settings which we believe deserve some special attention as part of future work.

\smallskip \noindent \textbf{Dynamic facility location}. Most of the work on the problem assumes a static model, in which the agents' locations are fixed over time. Exceptions to this are two fairly recent works \cite{de2018facility,fotakis2021reallocating} who considered multi-stage facility {\em reallocation} problems on the real line, where the facilities are being moved between stages based
on the locations reported by the agents and the goal is to minimize the social cost by keeping the reallocation cost low as well. Another exception is the work of Wada \emph{et al.} \cite{wada2018facility}, who
investigated settings with variable and dynamic populations, affecting the choices of the designer. We believe that there is ample space for future work here, with more dynamic settings potentially amenable to approximate mechanism design.

\smallskip \noindent \textbf{Incomplete information}. Another interesting avenue is to explore \emph{stochastic} settings, where the information about the agents' locations and preferences contains uncertainty, or is based on distributional assumptions. Along those lines, Caragiannis \emph{et al.} \cite{conf/icml/CaragiannisPS16}, motivated by machine-learning applications, considered single-facility location when agents' locations are independently and identically drawn from an unknown distribution. In a more ``possibilistic'' approach, Menon and Larson \cite{menon2019mechanism} considered a setting in which for each agent there is an interval on the line representing all her possible locations, and they explore robust mechanisms that perform well with respect to all the possible unknown true preferred locations of the agents within those intervals. These works have only scratched the surface of the effects of incomplete information on the design of mechanisms for the problem, and more work in this direction is needed.



\smallskip \noindent \textbf{Incentives beyond strategyproofness.} While strategyproofness is very desirable, it certainly impairs any attempt at good approximations, especially in settings where there are prohibitive lower bounds (e.g., the $n-2$ lower bound on the social cost for the two-facility location case). It is interesting to see if relaxing strategyproofness to an $\varepsilon$-approximate version, in which agents have \emph{very limited} incentive to misreport could ``open up'' the design space for much improved mechanisms.
For example, Sui  and  Boutilier \cite{sui2015approximately} provided several possibility/impossibility results for the $\varepsilon$-approximate (group) strategy-proofness in both constrained and unconstrained problems. Oomine \emph{et al.}
\cite{oomineSN17} studied the tradeoff between relaxation on group strategy-proofness and social benefit gain for obnoxious facilities.

 Another option would be to not require any form of strategyproofness and allow the agents to engage in a \emph{strategic game}; these games could be then studied using known techniques for analyzing \emph{Nash equilibria}, and their performance via the celebrated notion of the \emph{Price of Anarchy} \cite{koutsoupias1999worst}.
For example, Hotelling games are proposed by Hotelling \cite{hotelling1990stability} to model the situations where companies compete for attracting customers by strategically locating their facilities, and the customers prefer closer facilities.
Recent works have been focusing on characterizing the equilibria and bounding the corresponding Price of Anarchy (see \cite{feldman2016variations,ben2019multiunit}).
In another dimension, Sui and Boutilier \cite{sui2015optimal} study the computational problem of group manipulation: given a mechanism, how can the agents form collision to manipulate the outcome, in order to minimize the social cost? When the mechanism is ``Quantile", they solve this problem  for single-facility and multi-dimensional spaces  by formulating as cone programs. The group manipulation for other types of mechanisms is also worthy being studied.

\section{Concluding Remarks} From the abundance of different variants and interesting results that we have presented in this survey, it is evident that facility location problems are one of the most fundamental and most important examples of approximate mechanism design, and that it is at the very center of a vibrant literature in the major venues in artificial intelligence. And yet, despite the marvelous achievements that the literature on the problem has witnessed during the last decade, the future is nothing short of exciting. We believe that this survey will serve as a solid point of reference for current and future researchers in their quest to understand the major challenges, identify the most intriguing questions, and ultimately advance the state of the art even further.

\section*{Acknowledgement}
Bo Li is partially supported by The Hong Kong Polytechnic University (Grant No. P0034420). Minming Li is partially supported by NSFC (Grant No. 11771365).

\bibliography{references_abbr}

\begin{thebibliography}{10}

\bibitem{ahmadi2017survey}
A.~Ahmadi-Javid, P.~Seyedi, and S.~S. Syam.
\newblock A survey of healthcare facility location.
\newblock {\em Comput. Oper. Res.}, 79:223--263, 2017.

\bibitem{alon2010strategyproof}
N.~Alon, M.~Feldman, A.~D. Procaccia, and M.~Tennenholtz.
\newblock Strategyproof approximation of the minimax on networks.
\newblock {\em Math. Oper. Res.}, 35(3):513--526, 2010.

\bibitem{anastasiadis2018heterogeneous}
E.~Anastasiadis and A.~Deligkas.
\newblock Heterogeneous facility location games.
\newblock In {\em AAMAS}, pages 623--631, 2018.

\bibitem{anshelevich2018approximating}
E.~Anshelevich, O.~Bhardwaj, E.~Elkind, J.~Postl, and P.~Skowron.
\newblock Approximating optimal social choice under metric preferences.
\newblock {\em Artificial Intelligence}, 264:27--51, 2018.

\bibitem{archer2001truthful}
A.~Archer and E.~Tardos.
\newblock Truthful mechanisms for one-parameter agents.
\newblock In {\em FOCS}, pages 482--491, 2001.

\bibitem{aziz2020facility}
H.~Aziz, H.~Chan, B.~Lee, B.~Li, and T.~Walsh.
\newblock Facility location problem with capacity constraints: Algorithmic and
  mechanism design perspectives.
\newblock In {\em AAAI}, pages 1806--1813, 2020.

\bibitem{aziz2020capacity}
H.~Aziz, H.~Chan, B.~E Lee, and D.~C. Parkes.
\newblock The capacity constrained facility location problem.
\newblock {\em Games Econ. Behav.}, 124:478--490, 2020.

\bibitem{babaioff2012bitcoin}
M.~Babaioff, S.~Dobzinski, S.~Oren, and A.~Zohar.
\newblock On bitcoin and red balloons.
\newblock In {\em Proceedings of the 13th ACM conference on electronic
  commerce}, pages 56--73, 2012.

\bibitem{babaioff2016mechanism}
M.~Babaioff, M.~Feldman, and M.~Tennenholtz.
\newblock Mechanism design with strategic mediators.
\newblock {\em ACM Trans. Econ. Comput.}, 4(2):1--48, 2016.

\bibitem{ben2019multiunit}
O.~Ben-Porat and M.~Tennenholtz.
\newblock Multiunit facility location games.
\newblock {\em Math. of Oper. Res.}, 44(3):865--889, 2019.

\bibitem{black1948rationale}
D.~Black.
\newblock On the rationale of group decision-making.
\newblock {\em J. Polit. Econ.}, 56(1):23--34, 1948.

\bibitem{bordes1990strategy}
G.~Bordes, G.~Laffond, and M.~Le~Breton.
\newblock Strategy-proofness issues in some economic and political domains.
\newblock {\em Unpublished Manuscript, University of Bordeaux}, 1990.

\bibitem{brandt2016handbook}
F.~Brandt, V.~Conitzer, U.~Endriss, J.e Lang, and A.~D. Procaccia.
\newblock {\em Handbook of computational social choice}.
\newblock Cambridge University Press, 2016.

\bibitem{DBLP:conf/ijcai/CaiFT16}
Q.~Cai, A.~Filos{-}Ratsikas, and P.~Tang.
\newblock Facility location with minimax envy.
\newblock In {\em IJCAI}, pages 137--143, 2016.

\bibitem{sigecom/CaragiannisESY12}
I.~Caragiannis, E.~Elkind, M.~Szegedy, and L.~Yu.
\newblock Mechanism design: from partial to probabilistic verification.
\newblock In {\em {EC}}, pages 266--283, 2012.

\bibitem{conf/icml/CaragiannisPS16}
I.~Caragiannis, A.~D. Procaccia, and N.~Shah.
\newblock Truthful univariate estimators.
\newblock In {\em {ICML}}, pages 127--135, 2016.

\bibitem{tang11059mechanism}
X.~Chen, X.~Hu, X.~Jia, M.~Li, Z.~Tang, and C.~Wang.
\newblock Mechanism design for two-opposite-facility location games with
  penalties on distance.
\newblock In {\em SAGT}, pages 256--260, 2018.

\bibitem{chen2021tight}
X.~Chen, X.~Hu, Z.~Tang, and C.~Wang.
\newblock Tight efficiency lower bounds for strategy-proof mechanisms in
  two-opposite-facility location game.
\newblock {\em Inf. Process. Lett.}, 168:106098, 2021.

\bibitem{chen2019truthful}
X.~Chen, M.~Li, C.J. Wang, C~Wang, and Y.~Zhao.
\newblock Truthful mechanisms for location games of dual-role facilities.
\newblock In {\em AAMAS}, pages 1470--1478, 2019.

\bibitem{cheng2013strategy}
Y.~Cheng, W.~Yu, and G.~Zhang.
\newblock Strategy-proof approximation mechanisms for an obnoxious facility
  game on networks.
\newblock {\em Theor. Comput. Sci.}, 497:154--163, 2013.

\bibitem{cheng2015survey}
Y.~Cheng and S.~Zhou.
\newblock A survey on approximation mechanism design without money for facility
  games.
\newblock In {\em Advances in global optimization}, pages 117--128. Springer,
  2015.

\bibitem{church1978locating}
R.~L. Church and R.~S. Garfinkel.
\newblock Locating an obnoxious facility on a network.
\newblock {\em Transp. Sci.}, 12(2):107--118, 1978.

\bibitem{de2018facility}
B.~De~Keijzer and D.~Wojtczak.
\newblock Facility reallocation on the line.
\newblock In {\em IJCAI}, pages 188--194, 2018.

\bibitem{deligkas2021heterogeneous}
A.~Deligkas, A.~Filos-Ratsikas, and A.~Voudouris.
\newblock Heterogeneous facility location with limited resources.
\newblock {\em arXiv preprint arXiv:2105.02712}, 2021.

\bibitem{DingLCFN20}
Y.~Ding, W.~Liu, X.~Chen, Q.~Fang, and Q.~Nong.
\newblock Facility location game with envy ratio.
\newblock {\em Comput. Ind. Eng.}, 148:106710, 2020.

\bibitem{dokow2012mechanism}
E.~Dokow, M.~Feldman, R.~Meir, and I.~Nehama.
\newblock Mechanism design on discrete lines and cycles.
\newblock In {\em EC}, pages 423--440, 2012.

\bibitem{DBLP:conf/aldt/EscoffierGTPS11}
B.~Escoffier, L.~Gourv{\`{e}}s, N.~K. Thang, F.~Pascual, and O.~Spanjaard.
\newblock Strategy-proof mechanisms for facility location games with many
  facilities.
\newblock In {\em ADT}, pages 67--81, 2011.

\bibitem{feigenbaum2015strategyproof}
I.~Feigenbaum and J.~Sethuraman.
\newblock Strategyproof mechanisms for one-dimensional hybrid and obnoxious
  facility location models.
\newblock In {\em AAAI Workshops}, 2015.

\bibitem{Feigenbaum2014}
I.~Feigenbaum, J.~Sethuraman, and C.~Ye.
\newblock Approximately optimal mechanisms for strategyproof facility location:
  Minimizing $l_p$ norm of costs.
\newblock {\em Mathematics of Operations Research}, 42(2):434--447, 2017.

\bibitem{Distortion4}
M.~Feldman, A.~Fiat, and I.~Golomb.
\newblock On voting and facility location.
\newblock In {\em EC}, pages 269--286, 2016.

\bibitem{feldman2016variations}
M~Feldman, A~Fiat, and S.~Obraztsova.
\newblock Variations on the hotelling-downs model.
\newblock In {\em AAAI}, volume~30, 2016.

\bibitem{feldman2013strategyproof}
M.~Feldman and Y.~Wilf.
\newblock Strategyproof facility location and the least squares objective.
\newblock In {\em EC}, pages 873--890, 2013.

\bibitem{filimonov2021strategyproof}
A.~Filimonov and R.~Meir.
\newblock Strategyproof facility location mechanisms on discrete trees.
\newblock In {\em AAMAS}, pages 510--518, 2021.

\bibitem{DBLP:conf/aaai/Filos-RatsikasL15}
A.~Filos{-}Ratsikas, M.~Li, J.~Zhang, and Q.~Zhang.
\newblock Facility location with double-peaked preferences.
\newblock In {\em AAAI}, pages 893--899, 2015.

\bibitem{filos2020approximate}
A.~Filos-Ratsikas and A.~A. Voudouris.
\newblock Approximate mechanism design for distributed facility location.
\newblock {\em arXiv preprint arXiv:2007.06304}, 2020.

\bibitem{fong2018facility}
C.~K.~Ken Fong, M.~Li, P.~Lu, T.~Todo, and M.~Yokoo.
\newblock Facility location games with fractional preferences.
\newblock In {\em AAAI}, pages 1039--1046, 2018.

\bibitem{fotakis2021reallocating}
D.~Fotakis, L.~Kavouras, P.~Kostopanagiotis, P.~Lazos, S.~Skoulakis, and
  N.~Zarifis.
\newblock Reallocating multiple facilities on the line.
\newblock {\em Theoretical Computer Science}, 2021.

\bibitem{fotakis2010winner}
D.~Fotakis and C.~Tzamos.
\newblock Winner-imposing strategyproof mechanisms for multiple facility
  location games.
\newblock In {\em WINE}, pages 234--245, 2010.

\bibitem{fotakis2014power}
D.~Fotakis and C.~Tzamos.
\newblock On the power of deterministic mechanisms for facility location games.
\newblock {\em ACM Trans. Econ. Comput.}, 2(4):15:1--15:37, 2014.

\bibitem{fotakis2016strategyproof}
D.~Fotakis and C.~Tzamos.
\newblock Strategyproof facility location for concave cost functions.
\newblock {\em Algorithmica}, 76(1):143--167, 2016.

\bibitem{friedman2001elements}
J.~Friedman, T.r Hastie, and R.~Tibshirani.
\newblock {\em The elements of statistical learning}, volume~1.
\newblock Springer series in statistics New York, 2001.

\bibitem{fukui2020group}
Y.~Fukui, A.~Shurbevski, and H.~Nagamochi.
\newblock Group strategy-proof mechanisms for shuttle facility games.
\newblock {\em Journal of Information Processing}, 28:976--986, 2020.

\bibitem{golomb2017truthful}
I.~Golomb and C.~Tzamos.
\newblock Truthful facility location with additive errors.
\newblock {\em arXiv preprint arXiv:1701.00529}, 2017.

\bibitem{DBLP:conf/ijcai/GolowichNP18}
N.~Golowich, H.~Narasimhan, and David~C. Parkes.
\newblock Deep learning for multi-facility location mechanism design.
\newblock In {\em IJCAI}, pages 261--267, 2018.

\bibitem{hochbaum1982heuristics}
D.~S. Hochbaum.
\newblock Heuristics for the fixed cost median problem.
\newblock {\em Math. Program.}, 22(1):148--162, 1982.

\bibitem{hossain2020surprising}
S.~Hossain, E.~Micha, and N.~Shah.
\newblock The surprising power of hiding information in facility location.
\newblock In {\em AAAI}, pages 2168--2175, 2020.

\bibitem{hotelling1990stability}
H.~Hotelling.
\newblock Stability in competition.
\newblock In {\em The collected economics articles of Harold Hotelling}, pages
  50--63. Springer, 1990.

\bibitem{ibara2012characterizing}
K.~Ibara and H.~Nagamochi.
\newblock Characterizing mechanisms in obnoxious facility game.
\newblock In {\em COCOA}, pages 301--311, 2012.

\bibitem{koutsoupias1999worst}
E.~Koutsoupias and C.~Papadimitriou.
\newblock Worst-case equilibria.
\newblock In {\em STACS}, pages 404--413, 1999.

\bibitem{kyropoulou2019mechanism}
M.~Kyropoulou, C.~Ventre, and X.~Zhang.
\newblock Mechanism design for constrained heterogeneous facility location.
\newblock In {\em SAGT}, pages 63--76. Springer, 2019.

\bibitem{li2019strategyproof}
M.~Li, P.~Lu, Y.~Yao, and J.~Zhang.
\newblock Strategyproof mechanism for two heterogeneous facilities with
  constant approximation ratio.
\newblock In {\em IJCAI}, pages 238--245, 2019.

\bibitem{DBLP:conf/atal/LiMXZZ19}
M.~Li, L.~Mei, Y.~Xu, G.~Zhang, and Y.~Zhao.
\newblock Facility location games with externalities.
\newblock In {\em AAMAS}, pages 1443--1451, 2019.

\bibitem{li2020budgeted}
M.~Li, C.~Wang, and M.~Zhang.
\newblock Budgeted facility location games with strategic facilities.
\newblock In {\em IJCAI}, pages 400--406, 2020.

\bibitem{liu2021multiple}
W.~Liu, Y.~Ding, X.~Chen, Q.~Fang, and Q.~Nong.
\newblock Multiple facility location games with envy ratio.
\newblock {\em Theor. Comput. Sci.}, 2021.

\bibitem{lu2010asymptotically}
P.~Lu, X.~Sun, Y.~Wang, and Z.~A. Zhu.
\newblock Asymptotically optimal strategy-proof mechanisms for two-facility
  games.
\newblock In {\em EC}, pages 315--324, 2010.

\bibitem{lu2009tighter}
P.~Lu, Y.~Wang, and Y.~Zhou.
\newblock Tighter bounds for facility games.
\newblock In {\em WINE}, pages 137--148, 2009.

\bibitem{mei2019facility}
L.~Mei, M.~Li, D.~Ye, and G.~Zhang.
\newblock Facility location games with distinct desires.
\newblock {\em Discrete Applied Mathematics}, 264:148--160, 2019.

\bibitem{DBLP:conf/sagt/Meir19}
R.~Meir.
\newblock Strategyproof facility location for three agents on a circle.
\newblock In {\em SAGT}, pages 18--33, 2019.

\bibitem{menon2019mechanism}
V.~Menon and K.~Larson.
\newblock Mechanism design for locating a facility under partial information.
\newblock In {\em SAGT}, pages 49--62, 2019.

\bibitem{moulin1980strategy}
H.~Moulin.
\newblock On strategy-proofness and single peakedness.
\newblock {\em Public Choice}, 35(4):437--455, 1980.

\bibitem{DBLP:conf/ijcai/Narasimhan0P16}
H.~Narasimhan, S.~Agarwal, and D.~C. Parkes.
\newblock Automated mechanism design without money via machine learning.
\newblock In {\em IJCAI}, pages 433--439, 2016.

\bibitem{nehama2019manipulations}
I.~Nehama, T.~Todo, and M.~Yokoo.
\newblock Manipulations-resistant facility location mechanisms for zv-line
  graphs.
\newblock In {\em AAMAS}, pages 1452--1460, 2019.

\bibitem{NissimST12}
K.~Nissim, R.~Smorodinsky, and M.~Tennenholtz.
\newblock Approximately optimal mechanism design via differential privacy.
\newblock In {\em {ITCS}}, pages 203--213, 2012.

\bibitem{ono2017rename}
T.~Ono, T.~Todo, and M.~Yokoo.
\newblock Rename and false-name manipulations in discrete facility location
  with optional preferences.
\newblock In {\em PRIMA}, pages 163--179, 2017.

\bibitem{oomineSN17}
M.~Oomine, A.~Shurbevski, and H.~Nagamochi.
\newblock Parameterization of strategy-proof mechanisms in the obnoxious
  facility game.
\newblock {\em J. Graph Algorithms Appl.}, 21(3):247--263, 2017.

\bibitem{procaccia2009approximate}
A.~D. Procaccia and M.~Tennenholtz.
\newblock Approximate mechanism design without money.
\newblock In {\em EC}, pages 177--186, 2009.

\bibitem{procaccia2018approximation}
A.~D. Procaccia, D.~Wajc, and H.~Zhang.
\newblock Approximation-variance tradeoffs in facility location games.
\newblock In {\em AAAI}, 2018.

\bibitem{schummer2002strategy}
J.~Schummer and R.~V. Vohra.
\newblock Strategy-proof location on a network.
\newblock {\em J. Econ. Theory}, 104(2):405--428, 2002.

\bibitem{paolo2014heterogeneous}
P.~Serafino and C.~Ventre.
\newblock Heterogeneous facility location without money on the line.
\newblock In {\em ECAI}, pages 807--812, 2014.

\bibitem{paolo2015heterogeneous}
P.~Serafino and C.~Ventre.
\newblock Truthful mechanisms without money for non-utilitarian heterogeneous
  facility location.
\newblock In {\em AAAI}, pages 1029--1035, 2015.

\bibitem{sonoda2016false}
A.~Sonoda, T.~Todo, and M.~Yokoo.
\newblock False-name-proof locations of two facilities: Economic and
  algorithmic approaches.
\newblock In {\em AAAI}, volume~30, 2016.

\bibitem{sui2015approximately}
X.~Sui and C.~Boutilier.
\newblock Approximately strategy-proof mechanisms for (constrained) facility
  location.
\newblock In {\em AAMAS}, pages 605--613, 2015.

\bibitem{sui2015optimal}
X.~Sui and C.~Boutilier.
\newblock Optimal group manipulation in facility location problems.
\newblock In {\em ADT}, pages 505--520, 2015.

\bibitem{sui2013analysis}
X.~Sui, C.~Boutilier, and T.~Sandholm.
\newblock Analysis and optimization of multi-dimensional percentile mechanisms.
\newblock In {\em IJCAI}, pages 367--374, 2013.

\bibitem{tang2020characterization}
P.~Tang, D.~Yu, and S.~Zhao.
\newblock Characterization of group-strategyproof mechanisms for facility
  location in strictly convex space.
\newblock In {\em EC}, pages 133--157, 2020.

\bibitem{tang2020mechanism}
Z.~Tang, C.~Wang, M.~Zhang, and Y.~Zhao.
\newblock Mechanism design for facility location games with candidate
  locations.
\newblock In {\em COCOA}, pages 440--452. Springer, 2020.

\bibitem{DBLP:conf/atal/TodoIY11}
T.~Todo, A.~Iwasaki, and M.~Yokoo.
\newblock False-name-proof mechanism design without money.
\newblock In {\em AAMAS}, pages 651--658, 2011.

\bibitem{todo2020false}
T.~Todo, N.~Okada, and M.~Yokoo.
\newblock False-name-proof facility location on discrete structures.
\newblock In {\em ECAI}, pages 227--234, 2020.

\bibitem{wada2018facility}
Y.~Wada, T.~Ono, T.~Todo, and M.~Yokoo.
\newblock Facility location with variable and dynamic populations.
\newblock In {\em AAMAS}, pages 336--344, 2018.

\bibitem{walsh2020strategy3}
T.~Walsh.
\newblock Strategy proof mechanisms for facility location at limited locations.
\newblock {\em arXiv preprint arXiv:2009.07982}, 2020.

\bibitem{walsh2020strategy1}
T.~Walsh.
\newblock Strategy proof mechanisms for facility location in {Euclidean and
  Manhattan} space.
\newblock {\em arXiv preprint arXiv:2009.07983}, 2020.

\bibitem{journals/jair/Xu21Two}
X.~Xu, B.~Li, M.~Li, and L.~Duan.
\newblock Two-facility location games with minimum distance requirement.
\newblock {\em J. Artif. Intell. Res.}, 70, 2021.

\bibitem{xu2020strategyproof}
X.~Xu, M.~Li, and L.~Duan.
\newblock Strategyproof mechanisms for activity scheduling.
\newblock In {\em AAMAS}, pages 1539--1547, 2020.

\bibitem{yan2020strategyproof}
X.~Yan and Y.~Chen.
\newblock Strategyproof facility location mechanisms with richer action spaces.
\newblock {\em arXiv preprint arXiv:2002.07889}, 2020.

\bibitem{ye2015strategy}
D.~Ye, L.~Mei, and Y.~Zhang.
\newblock Strategy-proof mechanism for obnoxious facility location on a line.
\newblock In {\em COCOON}, pages 45--56, 2015.

\bibitem{yuan2016}
H.~Yuan, K.~Wang, C.~K.~K. Fong, Y.~Zhang, and M.~Li.
\newblock Facility location games with optional preference.
\newblock In {\em ECAI}, pages 1520--1527, 2016.

\bibitem{zhang2014strategyproof}
Q.~Zhang and M.~Li.
\newblock Strategyproof mechanism design for facility location games with
  weighted agents on a line.
\newblock {\em J. Comb. Optim.}, 28(4):756--773, 2014.

\bibitem{zou2015facility}
S.~Zou and M.~Li.
\newblock Facility location games with dual preference.
\newblock In {\em AAMAS}, pages 615--623, 2015.

\end{thebibliography}



\end{document}